\documentclass[10pt,journal,cspaper,compsoc]{IEEEtran}
\ifCLASSOPTIONcompsoc
\else
\fi
\ifCLASSINFOpdf
\else
\fi
\newcommand{\vnudge}{\vspace{-.1in}}
\newcommand{\mypara}[1]{{\bf{#1.} }}
\newcommand{\shortcite}[1]{\cite{#1}}

\usepackage[]{graphicx}
\usepackage{amsmath,amssymb}
\usepackage{algorithm2e}

\DeclareMathOperator*{\argmin}{arg\,min}

\begin{document}

%
\title{Ergonomic-driven \\ Geometric Exploration and Reshaping}
%
%
%
%

\author{Youyi~Zheng \quad\quad
        Julie~Dorsey \quad\quad
        Niloy~J.~Mitra
\IEEEcompsocitemizethanks{\IEEEcompsocthanksitem Y. Zheng and J. Dorsey are with the Department
of Computer Science, Yale University.\protect\\
E-mail: youyi.zheng@yale.edu; julie.dorsey@yale.edu
\IEEEcompsocthanksitem N. Mitra is with University College London.\protect\\
E-mail: n.mitra@cs.ucl.ac.uk}
\thanks{}}

%
%

\markboth{\emph{IEEE} Transactions on Visualization and Computer Graphics,~Vol.~xx, No.~xx}%
{Shell \MakeLowercase{\textit{et al.}}: Bare Demo of IEEEtran.cls for Computer Society Journals}
%


\IEEEcompsoctitleabstractindextext{%
\begin{abstract}
The paper addresses the following problem: given a set of man-made shapes, e.g., chairs, can we quickly rank and explore the set of shapes with respect to a given avatar pose? Answering this question requires identifying which shapes are more suitable for the defined avatar and pose; and moreover, to provide fast preview of how to alter the input geometry to better fit the deformed shapes to the given avatar pose? The problem naturally links physical proportions of human body and its interaction with object shapes in an attempt to connect ergonomics with shape geometry. We designed an interaction system that allows users to explore shape collections using the deformation of human characters while at the same time providing interactive previews of how to alter the shapes to better fit the user-specified character.
We achieve this by first mapping ergonomics guidelines into a set of simultaneous multi-part constraints based on target contacts; and then, proposing a novel contact-based deformation model to realize multi-contact constraints. We evaluate our framework on various chair models and validate the results via a small user study.
\end{abstract}

}

\maketitle

\IEEEdisplaynotcompsoctitleabstractindextext

%
\IEEEpeerreviewmaketitle

\section{Introduction}
\label{sec:intro}

\begin{figure}[b!]
  \centering
  \includegraphics[width=\linewidth]{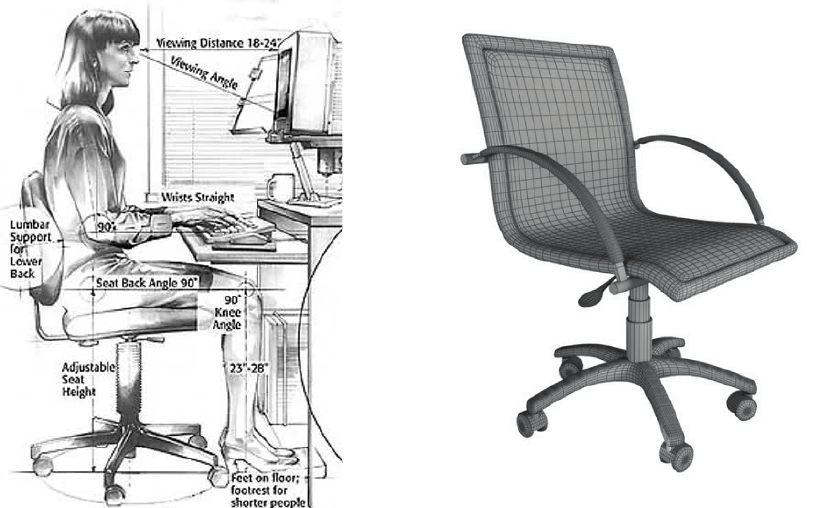}
  \caption{Ergonomic guidelines linking human posture to geometry of a workspace~(left). Traditionally, in computer graphics, raw model meshes~(right) remain oblivious of such specifications. }
  \label{fig:chair_specs}
\end{figure}

Humans come in various size and form.
The field of ergonomics focuses on accommodating such human variations with design goals of functional objects.
Various objects of daily use are shaped and given form based on their intended use
and target user. 
Guidelines on ergonomics (see \cite{book06} for a survey) summarize
years of such research, prototyping, product reviews, and design experiences to provide recommendations for geometric shapes based on their target usage (see Figure~\ref{fig:chair_specs}).

In computer graphics, object geometries are either acquired directly from or modeled inspired by
real objects. Hence, such objects often inherit or mimic real world object specifications.
In this work, we first investigate if one can classify and rank objects based directly on their target function
and associated ergonomic considerations. This is in contrast to typical classification strategies based on geometric descriptors and feature space analysis.
This leads to a novel {\em ergonomics-based} categorization, ordering, and exploration of input shape collections.

\begin{figure*}[t]
\centering
   \includegraphics[width=\linewidth]{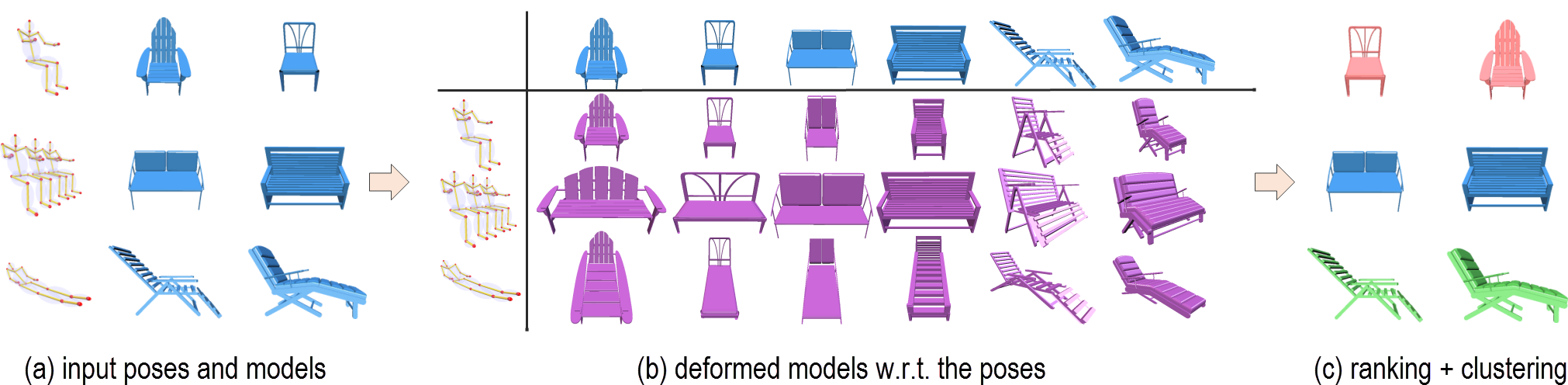}
   \caption{Starting from (a) a collection of shapes and a few user specified human avatar poses, (b) our method automatically reshapes the models to fit the avatars and ranks the models based on their deformation costs leading to a
    categorization of the input models driven by their compatibility with the avatar poses (c). }
   \label{fig:teaser}
   \vnudge
\end{figure*}

Moreover, evolution of digital fabrication has lead to feasible and economic custom design possibilities. Geometric
models can now be easily fabricated, making their target use relevant and important. Since robust solutions exist to
digitally capture high quality rigged digital avatars in matter of minutes (e.g.,~\cite{Tong:VirtualReality2012}),
as a second question, we investigate how to use such
human avatars specified in intended use poses to reshape geometry of virtual objects. This leads to a {\em ergonomics-driven} geometric reshaping of existing shapes.
By reshaping, we refer to adapting both the part proportions and their relative arrangements so the new shape better conforms to the target usage.
For example, Figure~\ref{fig:teaser} shows results of reshaping and classifying a set of chair models based on human avatar poses.

We address both questions using a novel contact-based deformation paradigm. First, we map ergonomic guidelines to a set of contact specifications between the human avatar
and input shape.
Such guidelines often come in the form of multiple specifications making it nearly impossible for lay users to satisfy by manual adjustment. For example, craftsman~\cite{chairFit:92} may make as many as twelve adjustments to the basic chair design (e.g., adjusting the headboard, lumbar support, or the angle between seat and back).
Starting from part-based input shapes, we first map such specifications to a set of multi-contact constraints.
We then propose a novel reshaping algorithm to computationally adapt an input shape to simultaneously satisfy these constraints.
%
We propose an iterative optimization procedure that alternately deforms the input shape and re-validates target contact specifications.

Our work makes the following two contributions: first, we link ergonomics considerations to shape reshaping with the goal to facilitate personal customization; second, we provide an ergonomics-driven shape exploration tool which subsequently benefits a novel shape classification setup. We evaluate our algorithm in the context of chairs, benches and sofas and validate the results through a small user study.


\if
ernomics for driving shape classification/editing
input: part model of objects + human avatar
we also know where a arm touches, etc. but *not* exact location for each part
so its a sliding contact, rather than an exact point-to-point
we first deform inputs to match target pose
and then (ii) rank the objects based on how close they are wrt poses
and (iii) fine scale ranking wrt how comfortable wrt candidate poses
(iv) we can edit the inputs based on different human size/proportion
collectively adapt 10 chairs for a particular person

iii) how to measure comfortable

An estimated 50\% of people in the industrialized world suffer some form of back complaint and many of these are related to poor seat design.

A common product chain in modern furniture design is to first conceptually sketch the product, followed by a tedious 3d modeling work flow and a set of complex and computational expensive physical simulations to validate the product. If the simulation process fails, the flow returns back to the design stage. Such procedure is typical in the industry production line for a new product and is rather expensive to conduct. As the number of 3d man-made shapes continues to grow, creating 3d models from sketches or imaginary now becomes a not-so-difficult task. However, the potential problem at the design stage remains, which is the fact the artists are not aware of or pay less attention to human factors (ergonomics) during their design process. Take the chair for an example, how we sit and what we sit on affects the health of the spine. A good chair should provide necessary support to the back, legs, buttocks, and arms, while reducing exposures to awkward postures, contact stress, and forceful exertions. A chair that is well-designed and appropriately adjusted is an essential element of a safe and productive computer workstation.

Increased adjustability ensures a better fit for the user, provides adequate support in a variety of sitting postures, and allows variability of sitting positions throughout the workday. This is particularly important if the chair has multiple users.
\fi

\section{Related Work}
Our work is closely related to existing works on shape exploration, geometric classification, and ergonomics study. While previous approaches exploit purely geometric or learning-based methods to achieve shape understanding, we focus on ergonomics both for classification and geometric deformation. 

\mypara{Shape exploration}
Recent research has proposed shape exploration tools to allow users to quickly browse large data collections. These methods either provide fast previews of shape variations \cite{Ovsjanikov11}, find partial correspondence \cite{Kim13}, or organize shapes into collective structures \cite{huang2013quartet}. The methods start from a pure geometric point of view to extract commonalities among a family of shapes using geometric descriptors or functional maps to find shape similarity. In contrast, we relate ergonomics with geometry to measure inter- and intra-class similarity for a given shape collection.

\mypara{Shape functional analysis}
Functionality of objects is closely related to their semantics. Understanding the shape functional properties remains the central challenge of existing shape analysis techniques. In recent years, a few methods have been proposed to reveal such connections based on purely geometry-based attempts \cite{laga_acmtog2013,Zheng:2014}. These approaches start from component-level models and build part-level contact graphs to facilitate subsequent analysis. In comparison, we exploit ergonomics guidelines to leverage a graph representation for the part-level deformation.

\mypara{Geometry and human factor}
Geometry reasoning has also been studied in computer vision and used for human workspace reasoning \cite{Fouhey12} or indoor scene understanding \cite{Gupta_CVPR11}. While these methods uses extensive training from existing semantic measures of image data to reason about a given particular image, the process of human measurement is typically performed offline. Coupling geometry modeling with physical simulation has been studied in recent research work of interactive chair modeling \cite{Saul:2010:SAC}. For an extensive discussion of history and development of
 ergonomics in the context of chairs we recommend Cranz's book~\cite{book88}, especially Chapters 3 and 5.

\mypara{Shape deformation}
Shape deformation has been a long-standing research topic in geometry processing \cite{DeformationSurvey:2008}. Existing shape deformation methods mainly fall into two classes: one class of methods aim to preserve the shape local properties such as curvatures, local coordinates, etc., \cite{Sorkine:2004,Lipman:2005:LRC,Lipman:2008:GC}, while the other class of methods aim to preserve global structures such as symmetry, inter-part relations \cite{Kraevoy:2008,Gal2009,ZFDOT10}.
Li et al.~\shortcite{Li_siga12}, in an interesting work, deforms input man-made objects to make them amendable to stacking.
In contrast, we focus on ergonomics guidelines for customizing chairs and sofas for specific target avatar sizes and poses. 

\begin{figure}[t!]
  \centering
  \includegraphics[width=\linewidth]{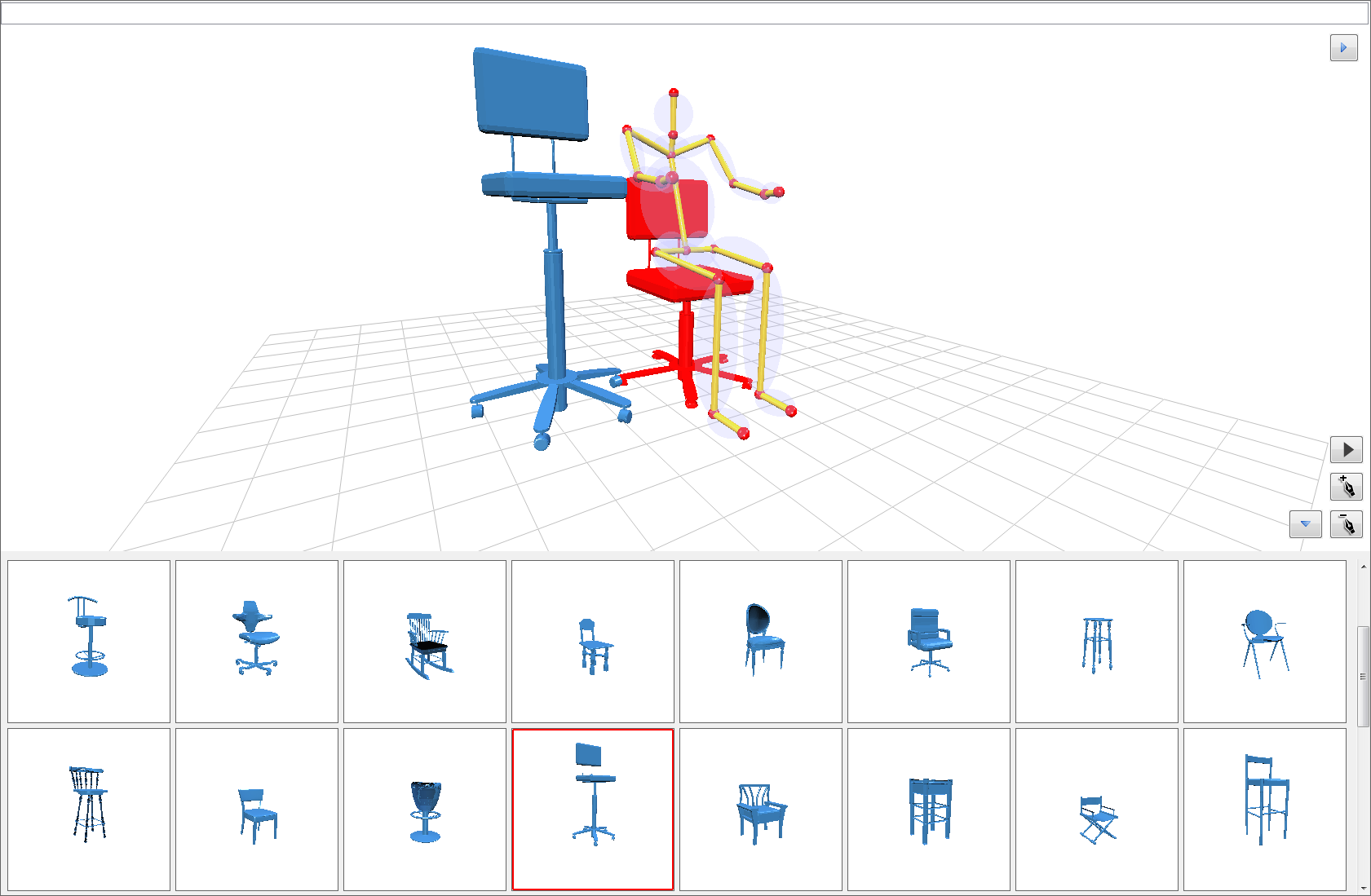}
  \caption{The user interface of our system consists of two panels: The top panel where users can interactively modify the human avatar and get intermediate feedback; and, the bottom panel to preview the ranked models according to the current avatar shape/pose.}
  \label{fig:user_interface}
  \vnudge
\end{figure}

\section {User Interface}
We now briefly describe our interactive framework (see Figure~\ref{fig:user_interface}). The interface consists of two panels: an interaction panel, which (i)~allows users to modify the human avatar shape, both its shape and its pose; (ii)~displays the original shape (in blue) and its deformed shape (in red) that fits the current human avatar; and a preview panel (bottom), which displays small windows of the ranked shapes according to how well they conform to the current avatar pose. The user can use the preview panel to browse through the  shapes. Once the user clicks a preview shape in the bottom panel, the shapes displayed in the top panel are updated to show the potential deformation needed to alter the shape to fit the current pose.

The user can directly modify the pose of the avatar by editing its skeleton nodes. The user can also refine the geometry by modifying the semantic attributes of the avatar such as leg length, body width,  hip width, etc. Each time the user finishes editing the character, the system interactively deforms the shapes and update the model rankings. We also allow the user to load pre-authored default poses. The supported avatar poses (for chairs) are: normal sitting, bench sitting, beach lying, and bar sitting (see Figure \ref{fig:avatar_pose}).

\section{Ergonomics Guidelines}
\label{sec:ergonomics}

Years of user studies and experience gathered from painful experience have been distilled in the form of qualitative ergonomics guidelines.
In this work we mainly focus on chairs. Although there is still some debate regarding the relative importance of the various guidelines, we summarize the ones that
are commonly suggested across the different works we consulted~\cite{book84,book88,chairGlove:92}:
\begin{itemize}
\item {\em Chair sears should have correct height} to allow both feet to be fully supported (e.g., by the ground). A chair that is too high creates undue pressure at the knee/thigh; while, if it is too short forces the knee to be higher
than the hip sockets.

\item {\em Width and depth of chair seats} should conform to the users dimensions. Specifically, while the width is dictated by the avatar's waistline, the depth is dictated by the length of the avatar's thigh bones.

\item {\em Flat uncontoured seats} are preferred to discourage a slouched or C-shaped posture.

\item {\em Lumbar support} by providing low- or mid-back support can help hold good posture and prevent pain to the spine and neck.

\item {\em Head support,} if provided, can help ease stress for the neck muscles and provide support for seating over extended periods.

\item {\em Arm rests} provide support for reading, typing, painting, and similar activities.

\end{itemize}

A seemingly obvious solution is to go for adjustable furniture, although at a higher cost. However, as observed by an early anthropometric study~\cite{2dof:63}, with more than two dimensions to (manually) adjust,
a person regularly forgets the previous (comfortable) setting among the large space of possible adjustments. Thus, paradoxically, with increased freedom, the users ends up adjusting their own posture to
fit an inconveniently dimensioned object.

\section{Algorithm Overview}
Once the user specifies a given avatar and an annotated pose (e.g., sitting, lying, bench sitting, etc), a set of geometric constraints are extracted by mapping ergonomics guidelines (see Section~\ref{sec:ergonomics}) to the underlying shape geometry, which are then integrated into the contact-based deformation paradigm to reshape the input model to conform to the avatar's pose.

\begin{figure}[t!]
\vnudge
  \centering
  \includegraphics[width=\linewidth]{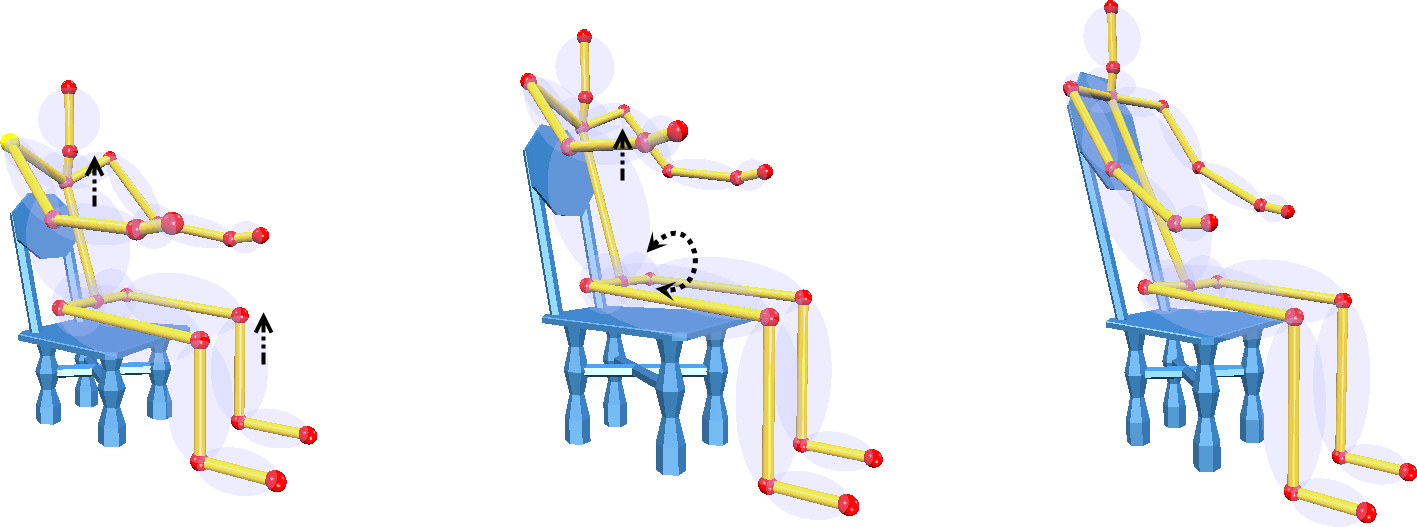}
  \caption{Modifying avatar's shape attributes (e.g., leg length, body length) will result in updates of the corresponding chair components, hence the styles.}
  \label{fig:deforming_specs}
\end{figure}

The reshaping works in two stages. In the analysis stage, the shape is represented by a spatial relation graph with nodes denoting its components and edges denoting the spatial relations (symmetry and contact) among the components. Given a specified avatar shape with an annotated pose, we first extract a set of ergonomic constraints based on the contact information between the character and the shape, we then design an edit propagation algorithm to deform the shape w.r.t. the ergonomic constraints while preserving its original structure. As the geometry and the pose of the character are altered, all constraints are automatically updated and a new deformed shape is generated. Thus, the user can design new shapes for a particular human character (e.g., design a beach chair for a kid). Figure \ref{fig:deforming_specs} shows that different avatars (in poses) lead to different deformed chairs.

Once the shape is deformed, we examine the deformation cost by measuring the shape volumetric and translational variations. This allows us to rank the shapes accordingly. The defined deformation cost is used for both shape ranking and shape classification. When classifying geometric content based on ergonomics, we start from a set of shapes and a few pre-defined human poses and compute the deformation cost for each chair to the given human shapes, and then cluster the shapes in the deformation space. By this, we can determine which chairs are more likely to be dinning chairs, or which chairs are more likely to be a beach chair, etc. Further, at a finer granularity, we also learn that within a class which shape is better suited for a particular avatar.

\begin{figure}[t!]
  \vnudge
  \centering
  \includegraphics[width=\linewidth]{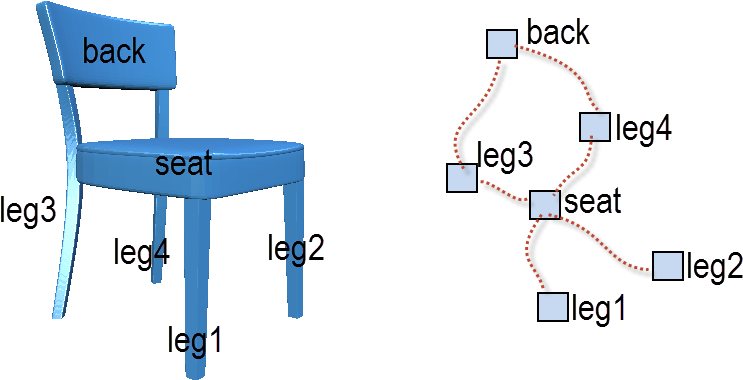}
  \caption{A chair encoded as a graph with nodes denoting the shape components while the edges representing the connection between two components (contact/symmetry).}
  \label{fig:representation}
\end{figure}

\section{The Method}


\subsection{Shape representation}
The input to our algorithm is a set of man-made objects that interact with our human body (e.g.,  chairs, sofas, etc.). We assume that the set of shapes are within the same category and are pre-aligned with multiple components that are consistently tagged (such as seat, back, arm, legs.) obtained via existing co-segmentation methods \cite{Huang11,Wang12}. 

Given a shape with its components, we represent the shape as a \emph{spatial relation graph} \cite{Zheng:2013}. Each node in the graph denotes a component while each contact and symmetry relation will be represented with a graph edge (see Figure \ref{fig:representation}). We equip each component with a primitive (in our system cuboid and cylinder, computed via PCA \cite{ZFDOT10}), we call such a primitive \emph{proxy}, which is later used for guiding the deformation of the underlying component. Once the shape components are associated with a set of proxies, we compute the contact information between adjacent components \cite{Kalogerakis12}.

In order to retain structure, what need to be preserved are the spatial relations among the components (or simply the spatial relations among proxies) and the individual shape properties of the underlying components. We show in next sections how such a representation enables a simple and robust contact-based deformation paradigm.

\begin{figure}[t!]
  \centering
  \includegraphics[width=\linewidth]{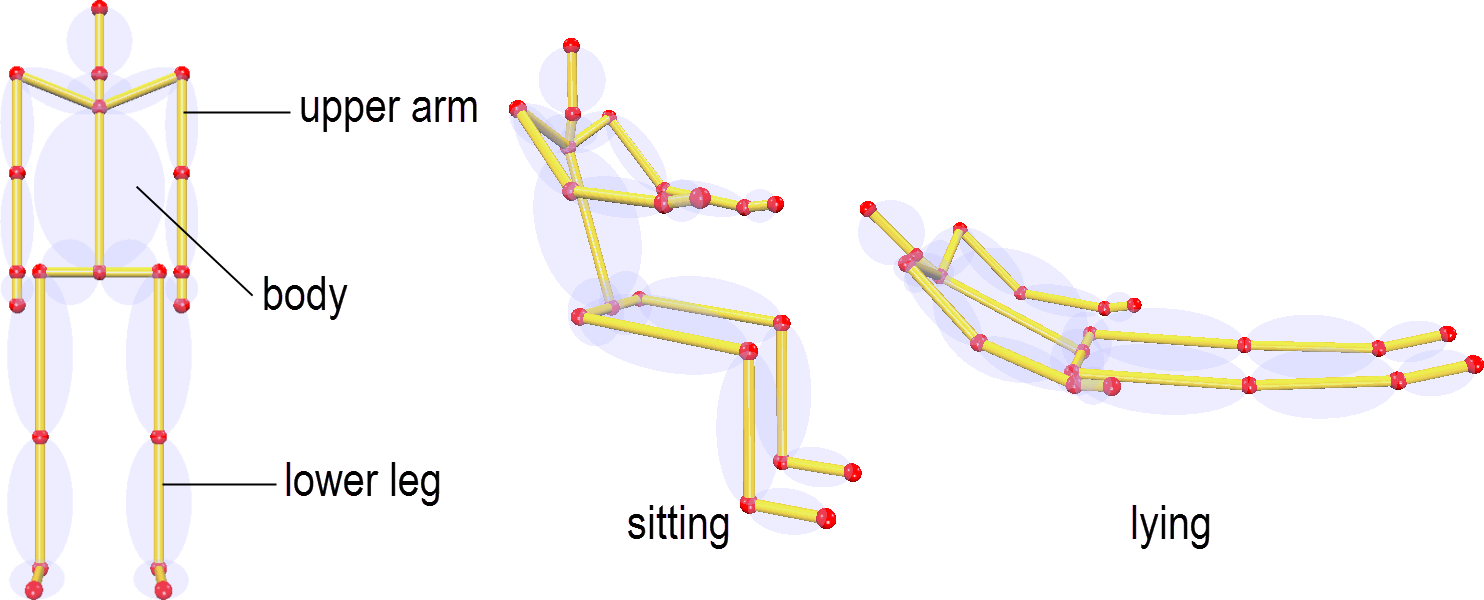}
  \caption{Human avatar used in our system consists of a skeleton whose bones representing body parts. The user can pose the avatar by dragging the skeleton nodes or change its shape by modifying semantic attributes such as leg length, body width, etc. The middle and right column show two representative poses used in our system.}
  \label{fig:avatar_pose}
  \vnudge
\end{figure}

\begin{figure}[b!]
  \vnudge
  \centering
  \includegraphics[width=\linewidth]{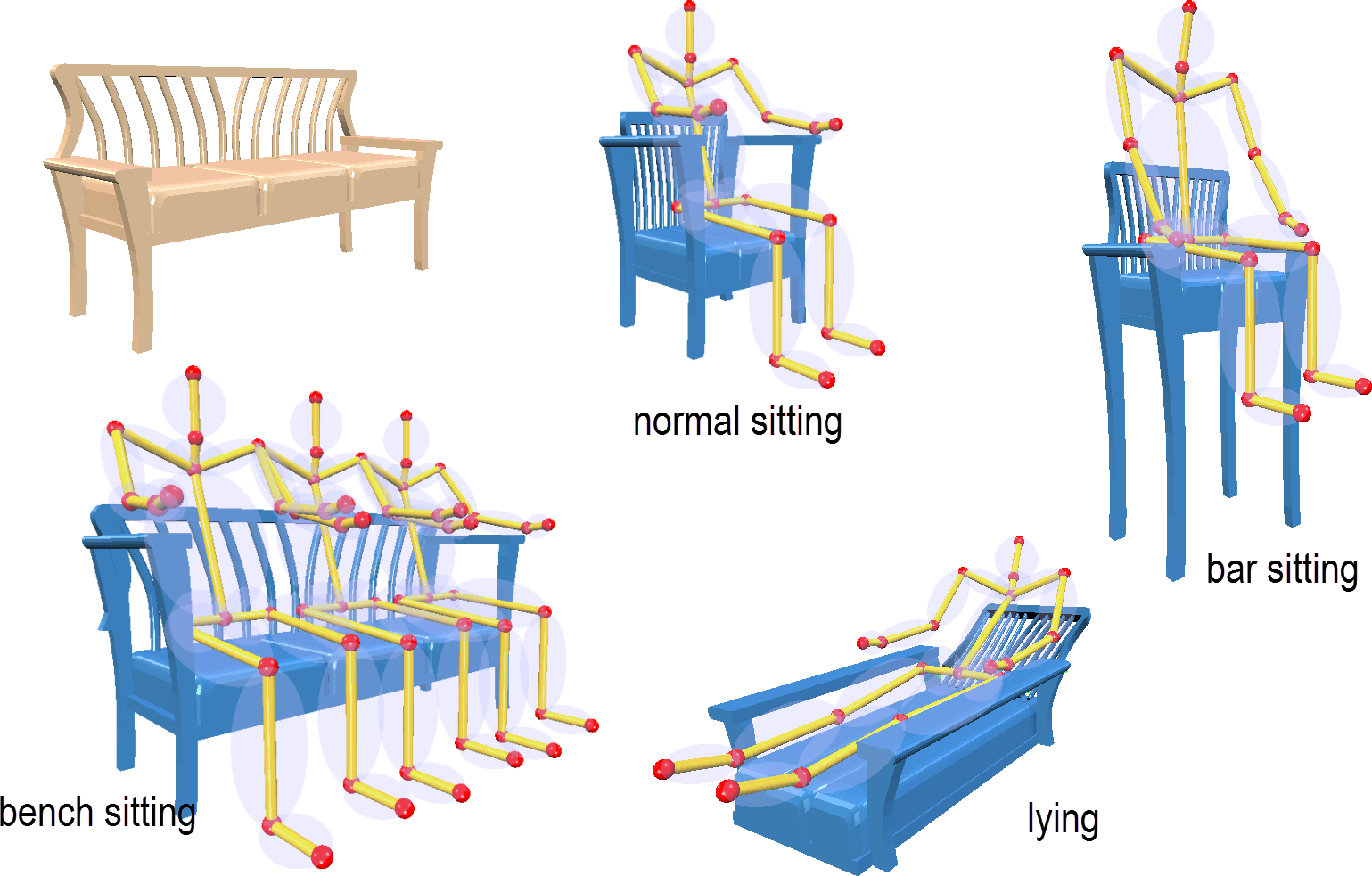}
  \caption{In our system, we design four types of avatar poses (sitting, lying, multiple avatars, bar sitting). Note that each pose corresponds to a chair style.}
  \label{fig:deforming_specs}
\end{figure}

\subsection{The human avatar}
Our system exposes to the user a predefined human character (Figure \ref{fig:avatar_pose}). We use a ellipsoid-based representation of the human body. The human skeleton is represented with a tree whose root node lies at the chest. Each skeleton bone is enclosed with an ellipsoid representing a body part. Each skeleton bone is also associated with attributes such as length, width, thickness, determining the dimensional properties of the ellipsoid for facilitating the user manipulation. In total there are 20 nodes and 19 bones.

\begin{figure*}[t!]
  \centering
  \includegraphics[width=\linewidth]{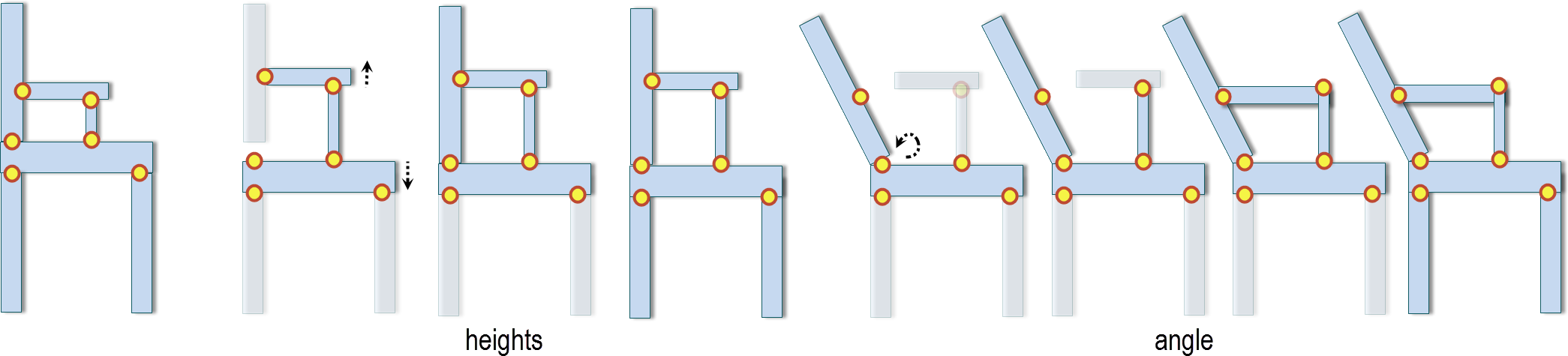}
  \caption{The contact-based deformation paradigm. Starting from a pre-sorted grouped ergonomic constraints, our deformation method iteratively visits the sets of constraints while conforming the shape to its original conductivities (i.e., contacts).}
  \label{fig:propagation}
  \vnudge
\end{figure*}

We annotate each bone and each node with semantic tags, for an instance, the skeleton bone corresponds to the body part is annotated as ``\emph{body-bone}". The user can use these semantic tags to alter the length, width, and thickness of individual body parts.

Besides the pre-specified poses, we allow the user to design poses. The user can simply drag a skeleton node, which is then projected into 3D space by mapping the translation from screen plane to the corresponding bone plane (defined by its consecutive skeleton bones). At this stage, all transformations are restricted to be rigid to preserve the body attributes. The user can simply move the entire body by dragging the root node (see accompanying video).

\subsection{Mapping ergonomic constraints}
We convert the ergonomics guidelines (Section~\ref{sec:ergonomics}) into geometric constraints to deform the shape and fit it to the human avatar. 
In the following, we will be mainly focus on chairs, while other examples are similarly handled.

The user first indicates the target pose, while our system provides potential suggestions of a given pose based on ergonomics guidance, e.g., the upper leg and lower leg should be orthogonal in a normal sitting pose while being almost parallel when seated on a beach chair.

\begin{figure}[b!]
  \vnudge
  \centering
  \includegraphics[width=\linewidth]{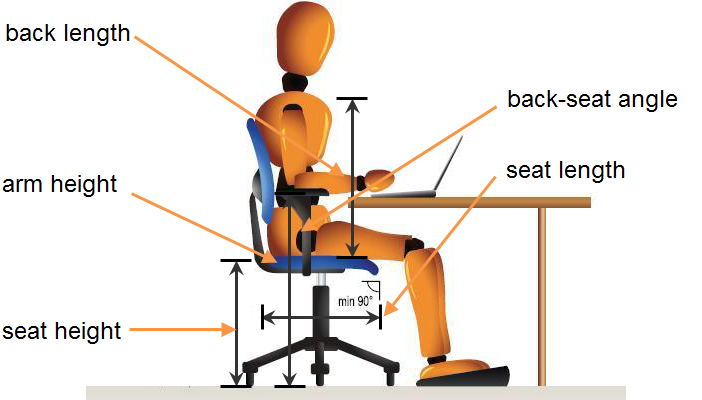}
  \caption{Mapping of ergonomic constraints to geometric constraints. Not all the constraints are shown.}
  \label{fig:mapping_ergonomics}
\end{figure}

Next, we identify a set of contacting regions on the body such as the hip should the top face of the seat, the lower arm will be in touch with chair arms, etc. All these types of ergonomic constraints are pre-specified into our system and recalled whenever a new pose is created by the user.

The contacting relations between the avatar and the chair are then computationally converted into geometric constraints. In particular, for chairs we design several types of constraints and name them based on the user annotated poses (normal sitting, lying, bar sitting, bench sitting, etc.). The constraints include:
\begin{itemize}
\item{seat width: determined by the hip width of the human body(s);}
\item{seat height: determined by the height of the hip;}
\item{seat length: determined by the length of the human upper legs;}
\item{arm height: determined by the position of the lower arms;}
\item{seat back angle: determined by the angle between the upper legs and the body spine, plus the supporting regions of the back and the hip;} and
\item{back length: determined by the body length and the supporting region.}
\end{itemize}
These ergonomic constraints can be directly discretized into geometric constraints in our system. For example, the contact relation between the hip and the chair seat will lead to a constraint in the height of the seat top face to be a specific value $h$. Other types of constraints, e.g., the angle between the body and the leg will lead to an sliding angle constraint between the seat and the back, and so on so forth. For those types of constraint, we allow for a $5\%-15\%$ range of sliding among the exact values to account for the subsequent deformation stiffness. For example, the seat width is allowed to be $1.1\times$--$1.2\times$ the width of the hips. Figure \ref{fig:mapping_ergonomics} shows an illustrative mapping of ergonomics to geometric constraints.

Since at this stage we are not dealing with actual physical fabrication but rather to explore, rate, and classify the shapes based on the human ergonomics as well as to have provide preview of deformations required to alter the given geometry, we do not enforce the constraints as hard ones. Note that all the geometric constraints (both length and angle) can be computationally approximated using contact constraints. Once these geometric constraints are derived, we attach them to the chair components and integrate into the contact-base deformation.

\begin{figure*}[t]
  \centering
  \includegraphics[width=\linewidth]{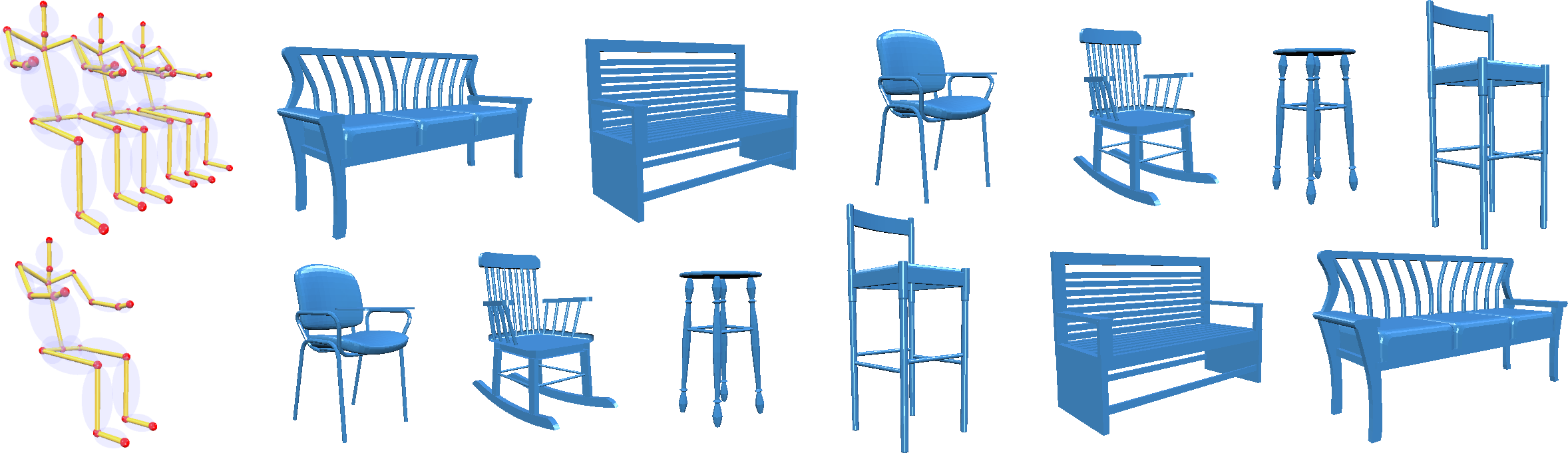}
  \caption{Browsing of chairs by specifying the avatar pose. Different poses lead to different ranking of the chairs. Here only 6 chairs out of 45 are shown.}
  \label{fig:chair_ranking}
\end{figure*}

\subsection{Contact-based deformation}
Given the set of constraints derived from human body and the chair, we now reshape the chair to meet these constraints while preserving the underlying shape structure. Previous efforts \cite{Gal2009,ZFDOT10} either use feature curves or controllers to induce the underlying deformation, which typically involves complicated strategies to delegate the edit among elements progressively or hierarchically. These methods proceed one edit at a time but cannot handle cases when multiple edit constraints are dependent with each other. 


Let us denote the set of components as $\{P_1, P_2, ..., P_n\}$ sorted by the number of ergonomic constraints associated with it and $\{C_1,C_2,...,C_k\}$ as groups of constraints by their names, i.e., heights, length, width, angle, etc. Since the constraints are not independent and can not be fulfilled in an initial setup (e.g., the angle between the back and seat will depends on the position of the seat), we greedily process them to enforce one type of constraints at a time, starting from $C_1 = \{c_1,c_2,...,c_l\}$ (e.g., the heights constraints). For each $c_i \in C_1$, we extract the transformation $T_i$ for the corresponding component so as to align it to meet $c_i$. For example, to lift a seat to a certain height $h$, the transformation is computed as a translation that maps the proxy center $o_i$ to a new position such that the height of the proxy's top face meet the height $h$. Width, length, and angle constraints are treated in a similar manner except that for those constraints which involve multiple components such as angle, for which we rotate and deform the proxies in accordance to the human body. Once all $c_i$-s $\in C_1$ has been handled, the deformation propagation starts. Transformation propagates from the already treated components to the rest untreated ones, based on contact and symmetry relations.

Figure \ref{fig:propagation} shows an overview of the contact-based propagation. Once we deform a component $P_i$, all its contacts (in terms of 3D points) are deformed accordingly (Figure \ref{fig:propagation} (b)). Let us denote the set of already treated components as $\Phi$ and the set of contacts as $\Theta$.  The propagation proceeds to one component at a time. Each time we look for one component which has the largest number of contacts link to others and denote it as $P_m$, then we look at the neighbors of $P_m$ in the relation graph. The one neighboring component $\chi$ which has not been treated and has the largest number of deformed contacts $\in \Theta$ will be selected as the next component to proceed (Figure \ref{fig:propagation} (c)). The chair seat in Figure \ref{fig:propagation} (c) is selected as $P_m$ since it has the largest number of contacts and the chair back component is selected as $\chi$ since it has the largest number of deformed contacts.

To deform a components w.r.t. its deformed set of contacts: $\{a_1,a_2,...\} \rightarrow \{a_1',a_2',...\}$, we find a best transformation matrix $T_{4\times{4}}$ such that the following energy is minimized:
\begin{equation}
T = \argmin_{T^*}\sum_i{\|T^*(a_i)-a_i'\|^2}.
\end{equation}
We solve the minimization in the least-squares senses. Enforcing the contact relations in a means retain the original spatial component structure. The propagation continues until all components are deformed. To reduce unnecessary deformations and preserve the original property of individual component, we adjust the $T$ as did in the method of \cite{ZFDOT10}, i.e., for a cylindrical shape, we retain its cylindrical property during deformation by enforcing uniform scales along its two non-principal axes. Figure \ref{fig:propagation} illustrates a simple 2{D} example of edit propagation.

Once $C_1$ is applied and the transformations are propagated, we proceed to $C_2$, $C_3$, and so on. As the propagation is invoked each time a $C_i$ is applied, there might be cases that when a $C_i$ is applied, it might alter the previous applied constraints $C_j$-s. To address this, we add an enforcing deformation term during the propagation, each time when a new component is to be treated, we align it with the previous constraints that are applied in previous $C_j$-s. Algorithm \ref{alg:propagation} is an overview of the contact-based deformation pipeline.

Note that unlike the propagation methods used in \cite{Gal2009,ZFDOT10} which process groups of elements (wires, controller feature curves) progressively, we focus on the contacts. We leverage the positions of contacts to guide the underlying component deformations where the enforcements of contacts naturally retain the spatial structure among shape components. Moreover, the typical number of contacts involved in a model is fairly small, which makes our method much simpler and faster than theirs. It also largely reduces the overhead that might be induced in interactive exploration stage once the collection size grows.

\begin{algorithm}[b!]
\vnudge
\caption{Contact-based deformation propagation. \vnudge }
 \SetAlgoLined
 \KwData{Input Model $\mathbf{M}:=\{P_1,\dots P_n\}$ and ergonomic constraint groups $\mathbf{C}:=\{C_1,\dots C_K\}$.}
 \KwResult{Updated model $M_i'$.}
 1. \While {$i < K$} {$C_i=\{c_1,c_2,...,c_l\} \in \mathbf{C}$\;
 a. \While {$j < l$} {
    i. apply $c_j \rightarrow P_r \in M$ \;
    ii. $\Phi \leftarrow P_r$ \;
  }
  \While {$\Phi \neq M$} {
    i. Find $P_m \in \Phi$ which has the maximum number of contacts and whose neighbors are not all in $\Phi$\;
    ii. Find a neighbor $P_q \in M/{\Phi}$ of $P_m$ which has the maximum number of deformed contacts\;
    iii. Deform($P_q$), w.r.t. its deformed contacts\;
    iii. $\Phi \leftarrow P_q$ \;
  }
}
\label{alg:propagation}
\end{algorithm}

\begin{figure}[t]
  \centering
  \includegraphics[width=\linewidth]{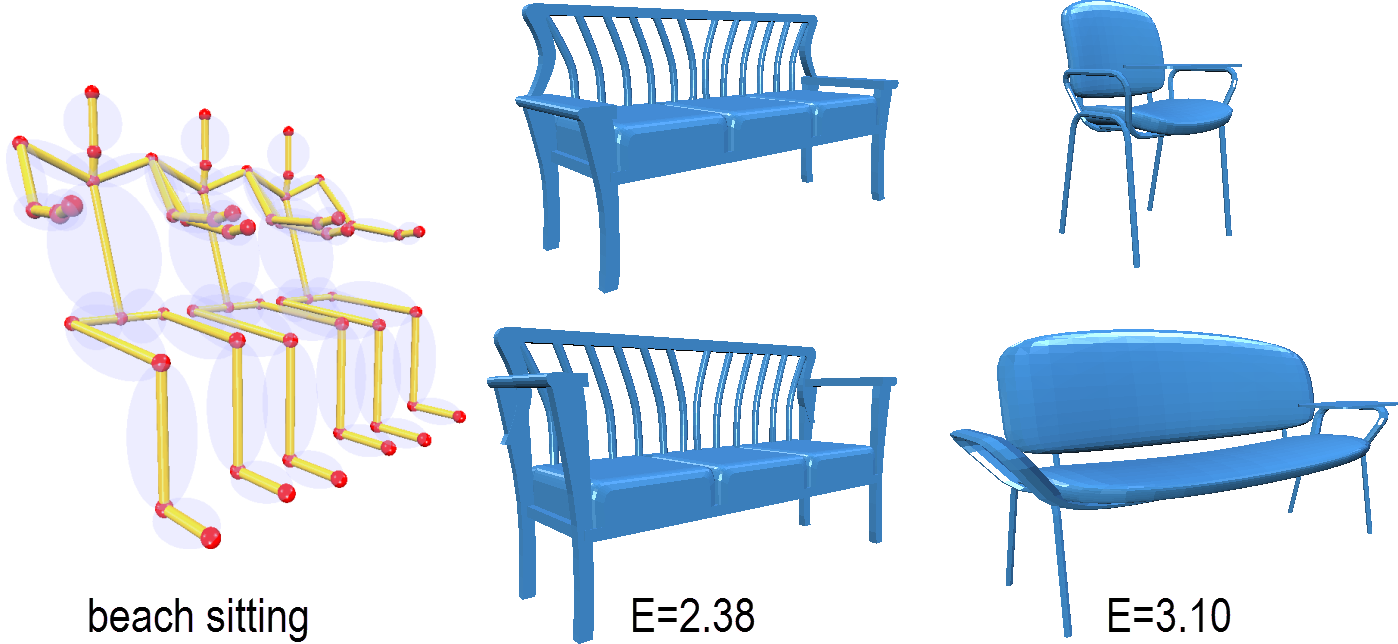}
  \caption{The deformation cost measured in terms of volumetric changes in three dimensions captures the velocity of valid transformation. It takes much more energy deforming an office chair into a bench chair.}
  \label{fig:chair_energy}
\end{figure}

\subsection{Geometric ranking}

Once a chair is deformed, we measure the deformation cost, i.e., how much energy it takes to deform the chair to fit the human avatar. By this setting, we can naturally rank the chairs according to the suitability to the current human avatar (and for browsing).

To measure the deformation cost of a chair model $M$, a straightforward method is to measure for each individual component how much deformation is required to deform it to its new shape. However, as the number of components, their sizes might vary across chairs, we resort to a more general computation. We first group the components into semantic parts based on their tags, such as back, seat, arm and base/legs. We then measure for each semantic part how much deformation is induced during the propagation by computing the scale changes in each dimension of their axis-aligned bounding boxes. For example, let us denote the bounding box of a chair part $i$ before deformation as $B_i$ and $B_i'$ after deformation. The deformation energy is measured as:
\begin{equation}
e_i = \Pi_{j=x,y,z}{|1+|\Delta{s_j}||} + \Pi_{j=x,y,z}{|1+|\Delta{t_j}||}.
\end{equation}
Here $\Delta{s_j}, \Delta{t_j}$ are the changes of scaling and translation in each dimension, i.e., $x,y,z$, of the bounding box of part $i$. The total deformation energy for $M$ is then defined as:
\begin{equation}
E_M = \frac{1}{N}\sum_{i=seat, back, ...}{e_i},
\end{equation}
where N is the number of semantic parts. Figure \ref{fig:chair_ranking} shows a simple ranking of different chairs w.r.t. the given avatar pose.

\begin{figure*}[t!]
  \centering
  \includegraphics[width=\linewidth]{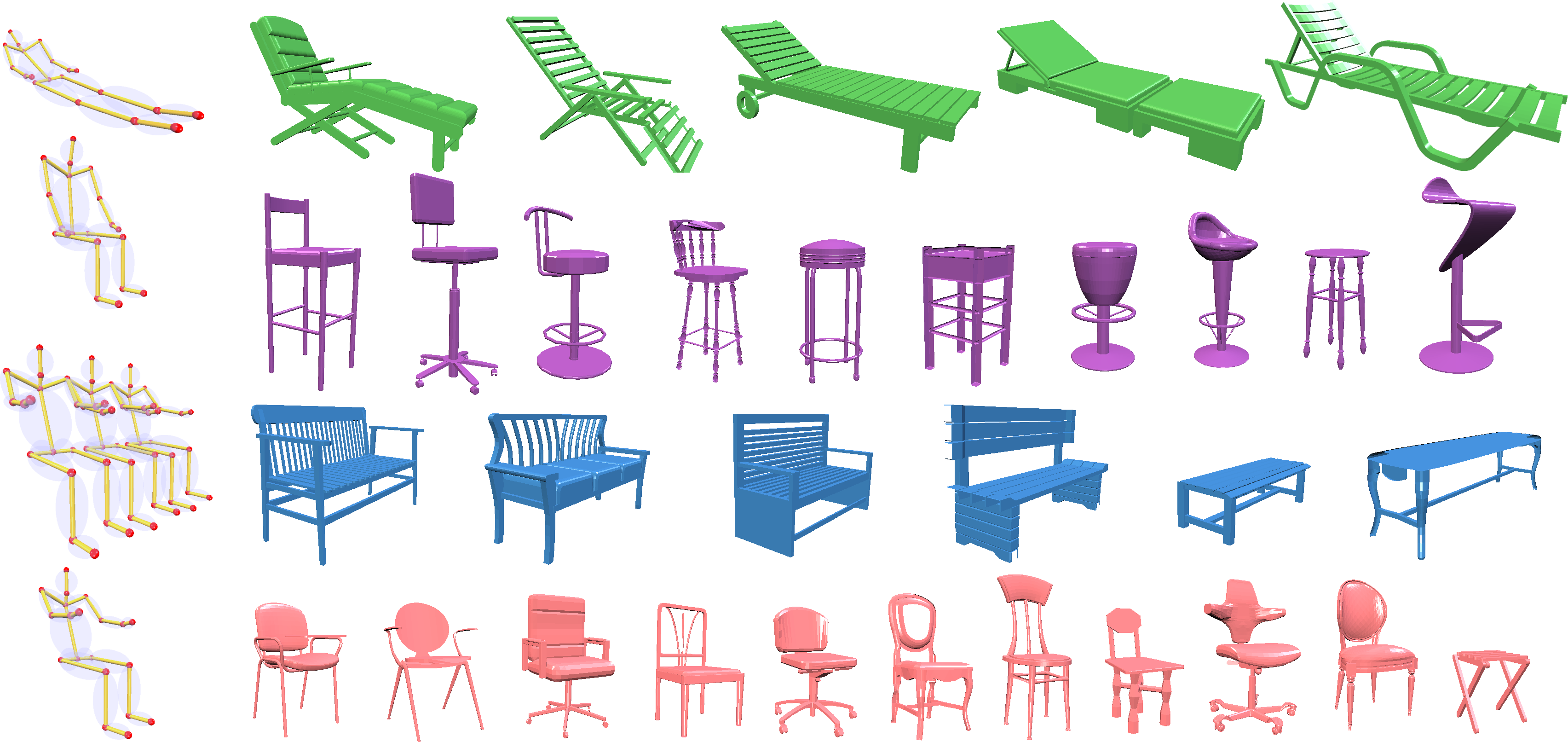}
  \caption{Clustering of a set of chairs w.r.t. 4 annotated poses: lying, bar sitting, bench sitting, and normal sitting. The color models show the clustering results. Note not all the models are shown.}
  \label{fig:chair_clustering}
  \vnudge
\end{figure*}

\section{Applications}

\subsection{Ergonomic-based Context classification}

A straightforward application of our method is ergonomic-based object classification. Given a set of chairs along with a few human annotated poses, we can identify which chairs are more likely to be sitting chair and which ones are more likely to be bench chairs, etc. The geometric classification is done by considering for each chair $M_i$, its deformation energy $E_{M_i}^k$ w.r.t. a particular pose $k$. Hence, given a set of predefined shapes, by comparing the corresponding energies, one can simply cluster the chairs w.r.t. the particular avatar poses. Figure \ref{fig:teaser}, \ref{fig:chair_clustering} show examples of chair classifications.

\subsection{Ergonomic-based Context co-retrieval}

\begin{figure}[b!]
  \vnudge
  \centering
  \includegraphics[width=\linewidth]{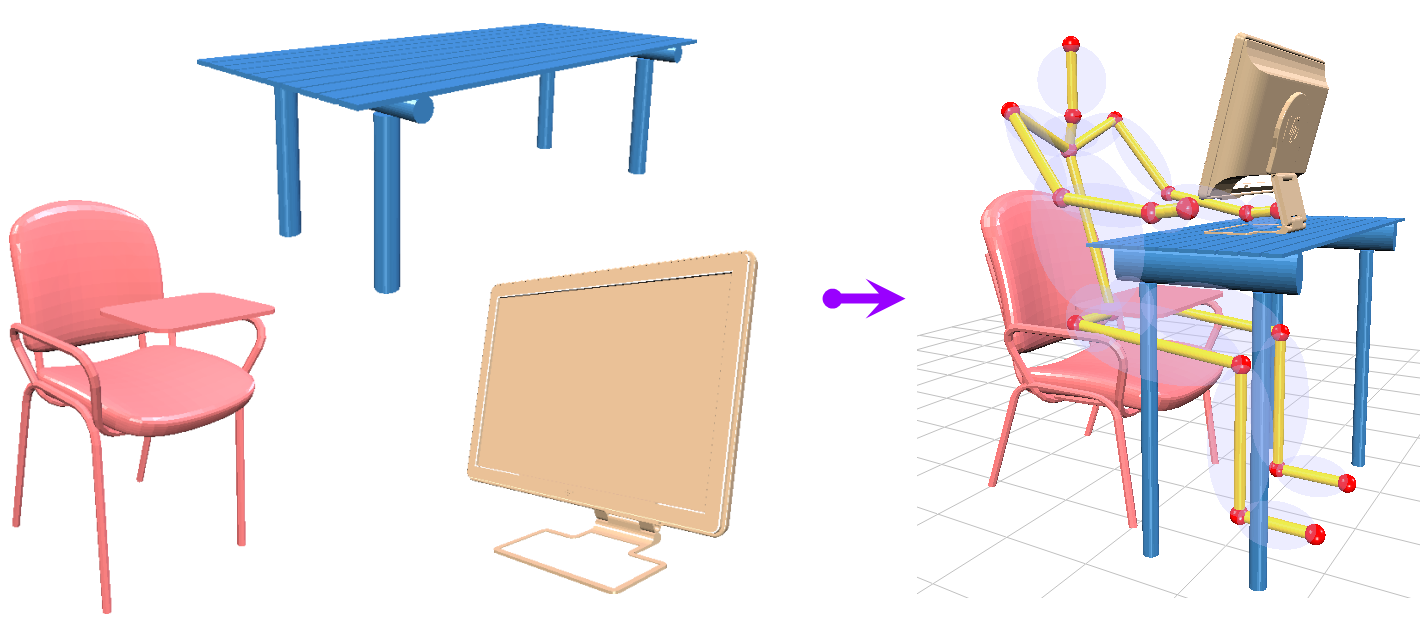}
  \caption{With our framework, we can co-retrieve multiple objects into a unified workplace using a single human avatar. In this example, the objects are retrieved in the order of chair, table, and monitor. Note that the size and the attributes of three objects are altered.}
  \label{fig:chair_coretrieval}
\end{figure}

Our human workspace is not designed separately for individual purposes. All man-made objects in a particular environment are co-related with particular people and human activities. At the central of the linkages lies the human. Our framework can be used for ergonomic-based objects co-retrieval. Given a specified human body shape, with user annotation, we can co-retrieve best suitable objects for a particular person based on ergonomics. For example, in Figure \ref{fig:chair_coretrieval}, the user annotated with a office sitting pose, and she then co-retrieves an office sitting chair, an office desk along with a monitor that is placed on the table. In this application, the human body plays the role to link the three objects with a conformation on ergonomics, sizes, and spatial relations. The ergonomic constraints are derived separately between the human body and the multiple objects while the placement of the three objects is taken into consideration also based on ergonomics (see Figure \ref{fig:chair_specs}). In this scenario, one is also able to navigate through multiple collections of shapes using a single human posed avatar for objects co-browsing in accordance to ergonomics. Note that in Figure \ref{fig:chair_coretrieval} the deformations are independently performed for the three objects.

\section{Evaluation}
We evaluate our algorithm mainly on man-made chair models (Figures \ref{fig:chair_ranking}, \ref{fig:chair_clustering}, and \ref{fig:chair_coretrieval}). We include a data set of 45 chairs consists of four main chair types: normal sitting chair/office chair, bench chair, beach chair, and bar chairs. Accordingly, we design four exemplar poses for these four chair types (Figure \ref{fig:avatar_pose}). For each pose, we extract a set of ergonomic constraints that are then attached to each chair during deformation. And for each pose, we evolve the chairs according these ergonomic constraints and rank them according to their deformation costs. In Figure \ref{fig:chair_ranking}, we show six different chairs with three different chair types, they are ranked according to the poses in the first column. The ranking clearly shows that which types of chairs are more suitable to the current posed human character.

We also evaluated the findings of the clustering algorithm with a user study. Specifically, we validated the classification results against a manually created ground truth. For each chair, we compute its deformation cost corresponding to each pose. Say, for a chair M, its deformation cost vector is $(E_M^1,E_M^2,\dots,E_M^k)$ where $k$ corresponds to the $k$-th pose. We then compute the pairwise similarity between each pair of chairs (defined as the L2 norm of their minimum components) and embed them into a 2{D} plot using MDS. We designed an browsing interface which allows the user to navigate through the clustered embedding (see also accompanying video). In our interface, the user can randomly hover the mouse onto the plotted 2{D} points, when the mouse pointer is close to a point, the corresponding chair model is shown in the main window. We let the user name the corresponding chair (office chair, bench chair, bar char, beach char, or none of the above). We measure the accuracy of the user named chairs against the ground truth (pre-tagged). We also asked 15 users to evaluate the results. Experiments showed that our algorithm achieves an average accuracy of over 95\% for each chair type. The main misleading cases are when a chair is not being selected as any of the given types.

\begin{figure}[t!]
  \centering
  \includegraphics[width=\linewidth]{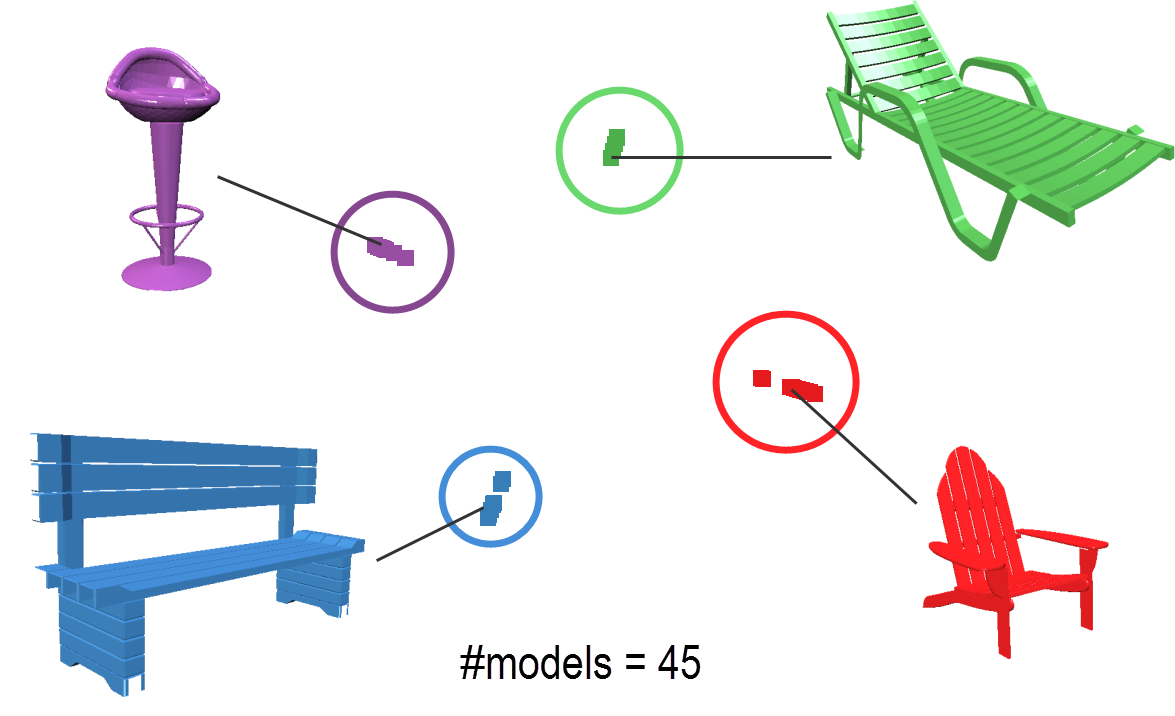}
  \caption{Embedding of chair models into 2D using multi-dimensional scaling~(MDS). Representative shapes are shown. Note different styles get clearly clustered.}
  \label{fig:chair_coretrival}
  \vnudge
\end{figure}

\textbf{Timing.} Our algorithm runs at interactive rate. The contact-based deformation paradigm is linear in terms of the number of components participated in the deformation. 
For a number of 45 chairs we tested, it took less than 2s to process them all (see also video). All experiments are done on a desktop with Intel i5-4430 processor (3.2GHZ) and 8GB memory.

\textbf{Limitations.} Our method has some limitations. First, our method leverages the ergonomics to deform the shapes to fit a given human character, the ergonomics derived from the human avatar and the underlying shape is not accurate but rough. Hence, our method can not be directly used for reshaping an object into a production-ready one. This can be an interesting future direction to explore. Second, our method assumes the input to have consistent multiple meaningful components (obtained via co-segmentation), if this assumption fails, users may need to manually segment and tag the input shapes.

\section{Conclusion}
In this paper, we introduce an algorithm to couple ergonomics with geometric exploration, rating, classification, and reshaping. The essential component of our algorithm is an ergonomics guided contact-based deformation paradigm that quickly evolves the shapes to fit the ergonomic constraints while allowing for fast preview of altered geometry to fit ergonomics and fast comparison among different shapes. 
We focus on concrete geometric aspects rather than abstracting them into a continuous space. Instead of browsing the shapes according to a given template, our method allows the user to deform a given human character in terms of shape and poses to explore the space of potential related shapes based on their suitability to the user character. Besides, our method provides realtime preview of how to alter the shape geometry to fit a given shape to the desired human character.

In the future, we further consider ergonomics for geometry analysis, potentially introducing physical simulators to examine fedility/stability of the object and also interaction with soft material (e.g., leather). We believe combining geometry with ergonomics can open up new opportunities towards the integrating functionality and usage with geometric modeling.



\ifCLASSOPTIONcaptionsoff
  \newpage
\fi



%

\bibliographystyle{IEEEtran}
\bibliography{geo_ergonomics}

\begin{thebibliography}{10}
\providecommand{\url}[1]{#1}
\csname url@samestyle\endcsname
\providecommand{\newblock}{\relax}
\providecommand{\bibinfo}[2]{#2}
\providecommand{\BIBentrySTDinterwordspacing}{\spaceskip=0pt\relax}
\providecommand{\BIBentryALTinterwordstretchfactor}{4}
\providecommand{\BIBentryALTinterwordspacing}{\spaceskip=\fontdimen2\font plus
\BIBentryALTinterwordstretchfactor\fontdimen3\font minus
  \fontdimen4\font\relax}
\providecommand{\BIBforeignlanguage}[2]{{%
\expandafter\ifx\csname l@#1\endcsname\relax
\typeout{** WARNING: IEEEtran.bst: No hyphenation pattern has been}%
\typeout{** loaded for the language `#1'. Using the pattern for}%
\typeout{** the default language instead.}%
\else
\language=\csname l@#1\endcsname
\fi
#2}}
\providecommand{\BIBdecl}{\relax}
\BIBdecl

\bibitem{book06}
G.~Salvendy, \emph{Handbook of Human Factors and Ergonomics}.\hskip 1em plus
  0.5em minus 0.4em\relax John Wiley \& Sons, Inc., 2006.

\bibitem{Tong:VirtualReality2012}
J.~Tong, J.~Zhou, L.~Liu, Z.~Pan, and H.~Yan, ``Scanning 3d full human bodies
  using kinects,'' \emph{Proc. IEEE Virtual Reality}, vol.~18, no.~4, pp.
  643--650, 2012.

\bibitem{chairFit:92}
R.~Erickson, ``Furniture by robert erickson: chairs that fit,'' brochure,
  Nevada city, 1992.

\bibitem{Ovsjanikov11}
\BIBentryALTinterwordspacing
M.~Ovsjanikov, W.~Li, L.~Guibas, and N.~J. Mitra, ``Exploration of continuous
  variability in collections of 3d shapes,'' \emph{ACM TOG (SIGGRAPH)},
  vol.~30, no.~4, pp. 33:1--33:10, 2011. [Online]. Available:
  \url{http://doi.acm.org/10.1145/2010324.1964928}
\BIBentrySTDinterwordspacing

\bibitem{Kim13}
V.~G. Kim, W.~Li, N.~J. Mitra, S.~Chaudhuri, S.~DiVerdi, and T.~Funkhouser,
  ``Learning part-based templates from large collections of 3d shapes,''
  \emph{ACM TOG (SIGGRAPH)}, vol.~32, no.~4, pp. 54:1--54:11, 2013.

\bibitem{huang2013quartet}
S.-S. Huang, A.~Shamir, C.-H. Shen, H.~Zhang, A.~Sheffer, S.-M. Hu, and
  D.~Cohen-Or, ``Qualitative organization of collections of shapes via quartet
  analysis,'' \emph{ACM Transactions on Graphics (Proceedings of SIGGRAPH
  2013)}, vol.~32, no.~4, p. accepted, 2013.

\bibitem{laga_acmtog2013}
H.~Laga, M.~Mortara, and M.~Spagnuolo, ``Geometry and context for semantic
  correspondence and functionality recognition in manmade 3d shapes,''
  \emph{ACM TOG}, p. to appear, 2013.

\bibitem{Zheng:2014}
Y.~Zheng, D.~Cohen-Or, M.~Averkiou, and N.~J. Mitra, ``Recurring part
  arrangements in shape collections,'' \emph{CGF (EUROGRAPHICS)}, vol.~33, p.
  to appear, 2014.

\bibitem{Fouhey12}
D.~F. Fouhey, V.~Delaitre, A.~Gupta, A.~A. Efros, I.~Laptev, and J.~Sivic,
  ``People watching: Human actions as a cue for single-view geometry,'' in
  \emph{Proc. 12th European Conference on Computer Vision}, 2012.

\bibitem{Gupta_CVPR11}
A.~Gupta, S.~Satkin, A.~A. Efros, and M.~Hebert, ``From 3d scene geometry to
  human workspace,'' in \emph{Computer Vision and Pattern Recognition(CVPR)},
  2011.

\bibitem{Saul:2010:SAC}
G.~Saul, M.~Lau, J.~Mitani, and T.~Igarashi, ``Sketchchair: An all-in-one chair
  design system for end users,'' in \emph{Proc. Tangible, Embedded, and
  Embodied Interaction}, 2011, pp. 73--80.

\bibitem{book88}
G.~Cranz, \emph{The Chair: Rethinking Culture, Body, and Design}.\hskip 1em
  plus 0.5em minus 0.4em\relax W. W. Norton \& Company, 1988.

\bibitem{DeformationSurvey:2008}
M.~Botsch and O.~Sorkine, ``On linear variational surface deformation
  methods,'' \emph{IEEE Transactions on Visualization and Computer Graphics},
  vol.~14, no.~1, pp. 213--230, 2008.

\bibitem{Sorkine:2004}
O.~Sorkine, D.~Cohen-Or, Y.~Lipman, M.~Alexa, C.~R\"{o}ssl, and H.-P. Seidel,
  ``Laplacian surface editing,'' in \emph{Proc. SGP}, 2004, pp. 175--184.

\bibitem{Lipman:2005:LRC}
\BIBentryALTinterwordspacing
Y.~Lipman, O.~Sorkine, D.~Levin, and D.~Cohen-Or, ``Linear rotation-invariant
  coordinates for meshes,'' in \emph{ACM SIGGRAPH 2005 Papers}, ser. SIGGRAPH
  '05.\hskip 1em plus 0.5em minus 0.4em\relax New York, NY, USA: ACM, 2005, pp.
  479--487. [Online]. Available:
  \url{http://doi.acm.org/10.1145/1186822.1073217}
\BIBentrySTDinterwordspacing

\bibitem{Lipman:2008:GC}
\BIBentryALTinterwordspacing
Y.~Lipman, D.~Levin, and D.~Cohen-Or, ``Green coordinates,'' \emph{ACM Trans.
  Graph.}, vol.~27, no.~3, pp. 78:1--78:10, Aug. 2008. [Online]. Available:
  \url{http://doi.acm.org/10.1145/1360612.1360677}
\BIBentrySTDinterwordspacing

\bibitem{Kraevoy:2008}
V.~Kraevoy, A.~Sheffer, A.~Shamir, and D.~Cohen-Or, ``Non-homogeneous resizing
  of complex models,'' \emph{ACM Trans. Graph.}, vol.~27, no.~5, pp.
  111:1--111:9, Dec. 2008.

\bibitem{Gal2009}
R.~Gal, O.~Sorkine, N.~J. Mitra, and D.~Cohen-Or, ``iwires: an analyze-and-edit
  approach to shape manipulation,'' \emph{ACM TOG (SIGGRAPH)}, vol.~28, no.~3,
  pp. 33:1--33:10, 2009.

\bibitem{ZFDOT10}
Y.~Zheng, H.~Fu, D.~Cohen-Or, O.~K.-C. Au, and C.-L. Tai, ``Component-wise
  controllers for structure-preserving shape manipulation,'' \emph{CGF
  (EUROGRAPHICS)}, vol.~30, no.~2, pp. 563--572, 2011.

\bibitem{Li_siga12}
H.~Li, I.~Alhashim, H.~Zhang, A.~Shamir, and D.~Cohen-Or, ``Stackabilization,''
  \emph{ACM TOG (SIGGRAPH Asia)}, vol.~31, no.~6, 2012.

\bibitem{book84}
A.~C. Mandal, \emph{The Sitting Position, Its Anatomy and Problems}.\hskip 1em
  plus 0.5em minus 0.4em\relax Daphne Publishing, 1984.

\bibitem{chairGlove:92}
G.~Gordon, ``Design a chair that fits like a glove,'' Fine Woodworking, 1992.

\bibitem{2dof:63}
F.~I.~R. Association, ``Anthropometric data: Limitation in use,'' Jour.
  Information Library, 1961.

\bibitem{Huang11}
\BIBentryALTinterwordspacing
Q.~Huang, V.~Koltun, and L.~Guibas, ``Joint shape segmentation with linear
  programming,'' \emph{ACM TOG (SIGGRAPH Asia)}, vol.~30, no.~6, pp.
  125:1--125:12, 2011. [Online]. Available:
  \url{http://doi.acm.org/10.1145/2070781.2024159}
\BIBentrySTDinterwordspacing

\bibitem{Wang12}
\BIBentryALTinterwordspacing
Y.~Wang, S.~Asafi, O.~van Kaick, H.~Zhang, D.~Cohen-Or, and B.~Chen, ``Active
  co-analysis of a set of shapes,'' \emph{ACM TOG (SIGGRAPH Asia)}, vol.~31,
  no.~6, pp. 165:1--165:10, 2012. [Online]. Available:
  \url{http://doi.acm.org/10.1145/2366145.2366184}
\BIBentrySTDinterwordspacing

\bibitem{Zheng:2013}
\BIBentryALTinterwordspacing
Y.~Zheng, D.~Cohen-Or, and N.~J. Mitra, ``{Smart Variations: Functional
  Substructures for Part Compatibility},'' \emph{CGF (EUROGRAPHICS)}, vol.~32,
  no. 2pt2, pp. 195--204, 2013. [Online]. Available:
  \url{http://dx.doi.org/10.1111/cgf.12039}
\BIBentrySTDinterwordspacing

\bibitem{Kalogerakis12}
E.~Kalogerakis, S.~Chaudhuri, D.~Koller, and V.~Koltun, ``A probabilistic model
  for component-based shape synthesis,'' \emph{ACM TOG (SIGGRAPH)}, vol.~31,
  no.~4, pp. 55:1--55:11, 2012.

\end{thebibliography}





%

\end{document}